\documentclass[preprint,12pt,authoryear]{elsarticle}

\usepackage{amssymb}
\usepackage{amsmath}

\usepackage{multirow}
\usepackage[switch,modulo]{lineno}
\usepackage{amsfonts}
\usepackage{amsthm}
\usepackage{mathrsfs}
\usepackage[title]{appendix}
\usepackage{xcolor}
\usepackage{textcomp}
\usepackage{manyfoot}
\usepackage{booktabs}
\usepackage{algorithm}
\usepackage{algorithmicx}
\usepackage{algpseudocode}
\usepackage{listings}
\usepackage{bm}
\usepackage{anyfontsize}
\usepackage{subcaption}
\usepackage{xurl}
\usepackage{hyperref}
\usepackage{orcidlink}
\begin{document}

\begin{frontmatter}



\title{The risks of dichotomising ordinal outcomes: A spatial analysis of self-rated health in Western Europe} 


\author[uv]{Miguel Ángel Beltrán-Sánchez}
\ead{angel.beltran@uv.es}

\author[uv]{Miguel Ángel Martínez-Beneito}
\ead{miguel.a.martinez@uv.es}

\author[ntnu-soc]{Terje Eikemo}
\ead{terje.eikemo@ntnu.no}

\author[ntnu-math]{Sara Martino}
\ead{sara.martino@ntnu.no}

\author[ntnu-math]{Andrea Riebler}
\ead{andrea.riebler@ntnu.no}

\affiliation[uv]{organization={Department of Statistics and Operational Research, University of Valencia},
            city={Burjassot},
            country={Spain}}

\affiliation[ntnu-soc]{organization={Department of Sociology and Political Science, Norwegian University of Science and Technology},
            city={Trondheim},
            country={Norway}}

\affiliation[ntnu-math]{organization={Department of Mathematical Sciences, Norwegian University of Science and Technology},
            city={Trondheim},
            country={Norway}}

\begin{abstract}
Survey responses are often measured using ordered response categories. In the European Social Survey, self-rated health is measured on a five-point scale from very good to very bad, yet analyses commonly dichotomise responses into binary categories of ``good'' and ``poor'' health. Binary indicators provide prevalence measures that are straightforward to communicate, but dichotomisation reduces information and may limit captured health variation, while the cut-off may influence estimates and substantive conclusions. Systematic evidence on these effects in spatial settings remains limited. We address this gap using European Social Survey round 11 (2023/2024) for Western Europe. We model self-rated health among male respondents by age, education and region. Bayesian spatial individual-level models with poststratification provide population-representative estimates. We compare an ordinal cumulative logit model for the five-category outcome with Bernoulli logistic regression models using two dichotomisations differing in the classification of ``fair'' health. Older age and lower education are consistently associated with worse self-rated health across specifications, suggesting relatively robust fundamental age and educational gradients. However, their magnitude and uncertainty, and some geographical conclusions, are sensitive to how the outcome is modelled. Assigning ``fair'' health to either side of a binary cut-off changes the regions identified as having above-average levels of less favourable health. The ordinal model retains category-specific information and can produce familiar binary prevalence estimates through aggregation. Binary indicators remain useful, particularly for monitoring and communication. Nevertheless, the selected cut-off should be justified and sensitivity to alternative cut-offs or an ordinal modelling approach should be considered, especially for geographical comparisons.
\end{abstract}




\end{frontmatter}



\section{Introduction} \label{sec:introduction}

In most large-scale surveys, respondents provide answers using ordered response categories. From a measurement perspective, these outcomes are therefore inherently ordinal. Nevertheless, a common practice in applied research is to dichotomise such measures into binary indicators \citep{Eikemo2008, Balaj2017}. While this simplification facilitates interpretation, allows the use of familiar logistic regression models and produces prevalence measures that are easy to communicate, the choice of cut-off is not always explicitly justified. Dichotomisation collapses adjacent categories and therefore reduces the information available from the original scale, as it treats responses on the same side of the cut-off as equivalent. As a result, it may attenuate gradients, mask heterogeneity and affect substantive conclusions. In addition, interpretations of model results may be particularly sensitive when a large central category is assigned to one or the other binary group.

The methodological literature has clearly highlighted potential limitations associated with the dichotomisation of ordinal or continuous variables \citep{Armstrong1989, MacCallum2002}. In particular, dichotomisation may reduce  the information contained in the original outcome measure, and statistical inference and substantive conclusions may be sensitive to the choice of threshold.

In social epidemiology and health-equity research, the use of dichotomised measures is standard and especially useful for communicating population health inequalities to policymakers. As such, dichotomisation remains widely used in spatial analyses of health inequalities based on survey data \citep{Thomson2017, Prag2017}. At the same time, there is still limited systematic empirical evidence on the extent to which dichotomisation affects spatial estimates, uncertainty and substantive interpretation. This gap is especially relevant because spatial health research increasingly relies on survey-based indicators, where modelling choices for ordinal outcomes can have direct implications for mapped patterns of health inequalities and potentially policy-relevant conclusions drawn from them.

Self-rated health (SRH) is a widely used indicator in sociology, social epidemiology and public health research to assess population health and its socio-economic patterning \citep{Jylha2009}. Despite its subjective nature, SRH has been widely validated and has consistently been shown to predict morbidity, health care use and mortality \citep{Idler1997, DeSalvo2006}. Its popularity stems from its ease of implementation in surveys: it is typically measured using a single question, such as ``How is your health in general?'', with ordered response categories. This makes it feasible to include in large surveys and easy to repeat over time. Because the wording is commonly standardised across major surveys, SRH also allows cross-survey and cross-national comparability. Moreover, it captures dimensions of health that are not easily observed in administrative data. Consequently, SRH plays a central role in monitoring health inequalities across socio-economic groups and geographical contexts. For example, \citet{Manor2000} used data from the 1958 British birth cohort to examine whether the relationship between socio-economic position and SRH, in terms of direction, magnitude or statistical significance, changed when SRH was analysed as a binary rather than an ordinal outcome. They found that the examined substantive conclusions were remarkably similar across modelling approaches. 

In this paper, we focus on the effect of dichotomisation in a spatial setting. We examine how different modelling strategies for SRH, binary versus ordinal, affect subnational prevalence estimates and the identification of geographical inequalities. Using data from round 11 of the European Social Survey (ESS11, 2023/2024) \citep{ESS2023}, we analyse SRH across Western Europe by age group, education level and NUTS region. Although rich administrative health data are available in some countries, they do not always include comparable socio-economic stratifiers such as education, making it difficult to study education-specific health inequalities using national registry sources alone. The ESS applies the same SRH question across countries and includes comparable information on age, education and region, making it possible to compare regions within a common measurement framework. ESS respondents evaluate their health using five ordered response categories: very good, good, fair, bad and very bad. 

We use Bayesian spatial individual-level models that account for dependence between neighbouring NUTS regions to compare an ordinal cumulative logit specification that preserves the five-category outcome with logistic regression models based on two commonly used dichotomisation schemes. These schemes differ in the classification of the central ``fair'' category, a modelling choice that may affect the resulting estimates and their interpretation. By comparing these modelling strategies and producing poststratified regional estimates, we provide empirical evidence on the practical implications of dichotomising ordinal health data in spatial inequality research. In particular, we assess which substantive conclusions remain consistent across specifications and which are sensitive to how the outcome is modelled. In doing so, the paper provides methodological guidance for researchers using SRH and contributes to improving the empirical study of health inequalities.

The paper is structured as follows. Section~\ref{sec:methods} introduces the Bayesian spatial individual-level models for binary-coded and ordinal SRH and describes poststratification as a statistical technique to obtain population-representative regional estimates. Section~\ref{sec:ess} introduces ESS11 and the Western European data used in the analysis. Section~\ref{sec:results} compares the results obtained from two alternative dichotomisation strategies with those obtained from the ordinal model. Section~\ref{sec:discussion} discusses the findings and their implications. 

\section{Methods} \label{sec:methods}

In this section, we introduce two Bayesian approaches for the spatial modelling of survey-based ordinal variables. The first approach, commonly used in practice, dichotomises the ordinal variable into a binary outcome through a selected cut-off point. The second approach preserves the ordered categories instead of collapsing them into a binary outcome. Both approaches are formulated at the individual level and are combined with poststratification, following the multilevel regression and poststratification approach \citep{Gelman1997, Park2004, Park2006}, to obtain finite population estimates for geographical areas and population subgroups.

Hereafter, we assume that the sample follows a stratified design, as is often the case in practice. This assumption allows us to focus on the modelling of the ordinal outcome rather than on the treatment of the survey sampling design, which is not the main focus of the paper. For simplicity, we present the model using a single categorical stratification variable. We denote the stratum of individual $i$ by $z_i \in \{1, \dots, Z\}$. In applications with several stratification variables, the same formulation can be extended by including them as separate main effects or, if appropriate, as a single combined factor.

Let $Y_i \in \{1, \dots, J\}$ denote an ordinal health indicator observed for $i = 1, \dots, n$ sampled individuals from a target population of size $N$, with categories ordered from better to worse health status. Let $r_i \in \{1, \dots, R\}$ denote the geographical unit, or region, to which individual $i$ belongs. We assume that population counts $N_{zr}$ are available for each poststratification cell defined by stratum $z$ and region $r$, so that $N$ is the sum of $N_{zr}$ over all strata and regions.

\subsection{Binary model based on a dichotomised version of ordinal variables} \label{subsec:binary}

A common modelling procedure is to replace the ordinal variable $Y_i$ by a binary recoding \citep{Armstrong1989, Ananth1997}. For a given cut-off point $c$, with $1 \leq c < J$, we define:
\begin{equation*}
Y_i^{D_c} =
\begin{cases}
0, & Y_i \leq c,\\
1, & Y_i > c,
\end{cases}
\label{eq:dichotomisation}
\end{equation*}
where $Y_i^{D_c}=1$ denotes membership of the less favourable health group under dichotomisation $D_c$. Different choices of $c$ may lead to different binary outcomes and, consequently, to different estimates and conclusions.

Direct stratified estimators, such as the Horvitz--Thompson estimator \citep{HT1952}, can be used to estimate finite population quantities by combining sample estimates within strata with the corresponding population proportions. For example, for a binary outcome in region $r$, such an estimator would combine the stratum-specific sample proportions using the weights $N_{zr}/N_r$, where $N_r$ is the population size of region $r$. In practice, direct estimates may also be obtained using the survey weights supplied with the data, particularly when the full sampling design or the relevant population counts are not available. The use of model-based approaches is motivated by the fact that direct stratified estimates can become unstable or even unavailable when the number of strata and regions is large relative to the sample size. In particular, some poststratification cells may contain few or no sampled respondents, which makes direct estimates infeasible. In this setting, individual-level models provide a way to estimate cell-specific probabilities by borrowing information across strata and neighbouring regions, thereby alleviating the problems posed by direct estimates.

For each dichotomisation $D_c$, we fit the following individual-level model for the binary outcome $Y_i^{D_c}$:
\begin{align*}
Y_i^{D_c} \mid \pi_{z_i r_i}^{D_c}
&\sim \text{Bernoulli}\left(\pi_{z_i r_i}^{D_c}\right),\\
\mathrm{logit}\left(\pi_{z_i r_i}^{D_c}\right)
&= \beta_0^{D_c} + \beta_{z_i}^{D_c} + \theta_{r_i}^{D_c}.
\label{eq:bernoulli_model}
\end{align*}
Here, $\pi_{zr}^{D_c}=P(Y_i^{D_c}=1 \mid z_i=z,r_i=r)$ is the probability of belonging to the less favourable health group for individuals in stratum $z$ and region $r$. The parameter $\beta_0^{D_c}$ is the intercept and represents the baseline log-odds for the reference stratum. The fixed effects $\beta_z^{D_c}$ measure differences in log-odds between stratum $z$ and the reference stratum, with $\beta_1^{D_c} = 0$ for identifiability. The term $\theta_r^{D_c}$ is a spatial random effect at the regional level, following a spatial distribution for areal data, such as BYM, Leroux or BYM2 \citep{Besag1991, Leroux2000, Riebler2016}. This term allows information to be shared between neighbouring regions and captures geographically structured residual variation after adjusting for the stratification variable. 

The model estimates cell-specific probabilities from the observed sample; these probabilities are then poststratified, as described in Section~\ref{subsec:poststratification}, to obtain population-representative estimates.

\subsection{Ordinal model for the original variable} \label{subsec:ordinal}

The binary model discards part of the information contained in the original ordinal variable $Y_i$. To retain this information, we instead model the ordered outcome directly. In particular, we consider the model proposed by \cite{BeltranSanchez2024} for the spatial analysis of ordinal survey-based variables. Let $\boldsymbol{\pi}_{z_i r_i} = \left(\pi_{1 z_i r_i}, \dots, \pi_{J z_i r_i}\right)^T$, where $\pi_{j z r} = P(Y_i = j \mid z_i = z, r_i = r)$ for $j = 1, \dots, J$. We assume the following likelihood:
\begin{equation*}
Y_i \mid \boldsymbol{\pi}_{z_i r_i} \sim \mathrm{Categorical}\left(\boldsymbol{\pi}_{z_i r_i}\right),
\label{eq:categorical_likelihood}
\end{equation*}
that is, $P(Y_i \mid \boldsymbol{\pi}_{z_i r_i}) = \pi_{1 z_i r_i}^{[Y_i = 1]}\cdot \ldots \cdot \pi_{J z_i r_i}^{[Y_i = J]}$, where $[Y_i=j]$ is equal to 1 if $Y_i=j$ and 0 otherwise.

For ordinal variables, it is natural to model the cumulative probabilities $\gamma_{j z r} = P(Y_i \leq j \mid z_i = z, r_i = r) = \sum_{k \leq j} \pi_{k z r}$, for $j = 1, \dots, J - 1$, rather than the category probabilities directly \citep{McCull1980, Congdon2005}. We therefore use a proportional odds cumulative logit model at the individual level:
\begin{equation*}
\mathrm{logit}(\gamma_{j z_i r_i}) = \kappa_j - \beta_{z_i} - \theta_{r_i}.
\label{eq:ordinal_model}
\end{equation*}
The cut-points (intercepts) $\boldsymbol{\kappa}$ satisfy $\kappa_1 < \dots < \kappa_{J - 1}$ and split the cumulative distribution on the logit scale. The fixed effects $\beta_z$ measure differences in cumulative log-odds between stratum $z$ and the reference stratum, with $\beta_1 = 0$ for identifiability. Under the proportional odds assumption, the effect of stratum $z$ is the same for all cumulative logits. This means that, relative to the reference stratum, each stratum is associated with more favourable or less favourable health responses across the ordered categories. The term $\theta_r$ is the region-level spatial random effect. We use this sign convention to make the interpretation of the effects parallel to that in the binary model. With the stratum and regional effects entering the cumulative logit predictor with a minus sign, positive values of $\beta_z$ or $\theta_r$ are associated with higher probabilities of less favourable health categories. The connection between the ordinal and dichotomised formulations is formalised in Supplementary Material~A.

Once the model has been fitted, estimates of the cumulative probabilities $\gamma_{jzr}$ can be obtained for each cut-point $j$, stratum $z$ and region $r$. Since the likelihood is written in terms of category probabilities, these are recovered from the cumulative probabilities as: $\pi_{1zr} = \gamma_{1zr}$, $\pi_{jzr} = \gamma_{jzr} - \gamma_{j-1,zr}$ for $j = 2, \dots, J - 1$, and $\pi_{Jzr} = 1 - \gamma_{J-1,zr}$. Therefore, the ordinal model avoids selecting a cut-off point and provides estimates for each category of the original outcome. In addition, any binary or grouped version of the outcome can be recovered from the fitted ordinal model by summing the corresponding category probabilities. This allows direct comparison between models fitted after dichotomisation and aggregations derived from the ordinal model.

Although the binary and ordinal models have been presented without specifying a particular inferential framework, both can be estimated from either a frequentist or a Bayesian perspective. In a Bayesian formulation, prior distributions are assigned to the model parameters, and inference is based on the resulting posterior distribution. In this paper, we follow a Bayesian approach for inference. This formulation is particularly convenient for the spatial component, as it allows spatial dependence between neighbouring regions to be incorporated directly through the prior distribution of the regional effects. It also provides a natural way to carry uncertainty from the model parameters to the poststratified population estimates. 

\subsection{Poststratification and finite population quantities} \label{subsec:poststratification}

Once the individual-level models have been fitted to the survey data, they provide probability estimates for each combination of stratum and region; that is, $\widehat{\pi}_{zr}^{D_c}$ for the binary model and $\widehat{\pi}_{jzr}$ for the ordinal model. These probabilities are model-based indirect estimates for the corresponding poststratification cells, instead of direct estimates based only on the respondents in each cell. Poststratification \citep{Little1993} combines these cell-specific probability estimates with the corresponding population counts, so that the resulting estimates reflect the composition of the target population rather than only the composition of the observed sample. Let $N_r=\sum_{z=1}^{Z} N_{zr}$ denote the population size of region $r$.

For a binary outcome defined by dichotomisation $D_c$, the finite population prevalence of the less favourable health group in region $r$, $P_r^{D_c}$, is estimated as:
\begin{equation*}
\widehat{P}_r^{D_c} = \frac{1}{N_r} \sum_{z = 1}^{Z} N_{zr} \widehat{\pi}_{zr}^{D_c}.
\label{eq:post_binary}
\end{equation*}
Equivalently, this is a weighted average of the model-based stratum-specific probabilities within region $r$, with weights given by the population distribution of the strata in that region. In this way, poststratification provides an estimate of the quantity of interest, at the population level, for region $r$.

For the ordinal model, poststratification is applied separately to each response category. The finite population proportion in category $j = 1, \dots, J$ and region $r$, $P_{jr}$, is estimated as:
\begin{equation*}
\widehat{P}_{jr} = \frac{1}{N_r} \sum_{z = 1}^{Z} N_{zr} \widehat{\pi}_{jzr}.
\label{eq:post_ordinal}
\end{equation*}
Thus, the ordinal model provides one poststratified estimate for each original response category. These estimates can also be aggregated to reproduce any binary or grouped version of the outcome. For any set of categories $A \subseteq \{1,\dots,J\}$, the corresponding poststratified regional proportion is:
\begin{equation*}
\widehat{P}_r^{A} = \sum_{j \in A} \widehat{P}_{jr}.
\label{eq:post_ordinal_aggregation}
\end{equation*}
This makes it possible, for the same dichotomisation, to compare a binary model fitted directly to the dichotomised outcome with the corresponding aggregation derived from the ordinal model.

Uncertainty in the poststratified quantities is obtained by applying the same poststratification expressions to each posterior draw of the cell-specific probabilities. The posterior means, standard deviations, coefficients of variation, prediction intervals and exceedance probabilities reported in the results are then calculated from the resulting posterior samples of the regional estimates. This propagates model uncertainty to the population-level estimates, while treating the population counts $N_{zr}$ as fixed.

In both the binary and ordinal settings, poststratification yields population-representative estimates that adjust for potential differences between the composition of the sample and that of the target population. This requires information on the population size of each stratum within each region, $N_{zr}$, obtained from census data, population registers, administrative records or other external sources. Although the focus of this paper is on regional estimates, the same poststratification procedure can be used to obtain estimates for other population subgroups, corresponding to particular combinations of categories of the stratification variables, or higher levels of aggregation, provided that the corresponding population counts are available.

\section{Self-rated health in Western Europe using ESS11} \label{sec:ess}

We used data from the 11th round of the European Social Survey (ESS11), corresponding to 2023/2024\footnote{\url{https://ess.sikt.no/en/datafile/242aaa39-3bbb-40f5-98bf-bfb1ce53d8ef}}. The ESS is a periodic cross-national survey designed to provide high-quality comparative data on social attitudes, behaviours and living conditions across Europe. Its sampling procedures are country-specific and cover each participating country's resident population aged 15 years or older \citep{ESS2023}. This makes the ESS a particularly useful source for studying population health and social inequalities in a comparative European setting \citep{Hoven2025}.

The outcome analysed in this paper is subjective general health, the label used in the ESS questionnaire. Throughout the paper, we refer to this variable as self-rated health (SRH), following the terminology commonly used in the health inequalities literature \citep{Idler1997, Jylha2009}. The question wording was: ``How is your health in general? Would you say it is...''. Response options were 1 = very good, 2 = good, 3 = fair, 4 = bad and 5 = very bad. 

The Western European sample used here includes Austria, Belgium, Switzerland, Germany, Spain, France, Italy, the Netherlands and Portugal. These countries were selected both because they provide a substantively meaningful Western European case study and because population counts by region, sex, age and education were available for poststratification. Population data were obtained from Eurostat Census 2021 tables\footnote{\url{https://ec.europa.eu/eurostat/databrowser/view/cens_21cobe_r2/default/table?lang=en&category=cens.cens_21.cens_21dc}}. These external population counts are required to obtain population-representative estimates through the poststratification procedure described in Section~\ref{subsec:poststratification}. The sample contains 17,053 respondents, including 8,097 male and 8,956 female respondents. The main analysis in this paper focuses on male respondents. The same analysis is applied to female respondents, and the corresponding results are reported in Supplementary Material~B.

The countries above are divided into 102 subnational regions. In ESS11, regional information is available at NUTS-2 level for Austria (9 regions), Belgium (11), Switzerland (7), Spain (16), France (21), the Netherlands (12) and Portugal (5), and at NUTS-1 level for Germany (16) and Italy (5).

The variables used in the analysis are:
\begin{itemize}
    \item \textit{Self-rated health}. The original five-category SRH variable is retained for the ordinal analysis and recoded using two alternative cut-offs for the binary analyses.
    \item \textit{Age}. Age is grouped into three categories: 15--34, 35--54 and 55 years or older.
    \item \textit{Education}. Educational attainment is classified according to the International Standard Classification of Education (ISCED) and grouped into three categories: ISCED $\leq 2$, $3 \leq \operatorname{ISCED} \leq 4$ and ISCED $\geq 5$.
    \item \textit{Region}. Region of residence is coded according to the $102$ NUTS regions described above.
\end{itemize}
The age and education groupings are consistent with those used in the ESS poststratification weighting procedure. In the ESS, poststratification weights use information on region, sex, age and education \citep{ESSWeighting}. However, the ESS sampling design differs across countries, and modelling all country-specific design features is not the focus of this paper. For the purposes of the model-based analysis, we assume that the sampling design is ignorable conditional on these variables.

For the analyses, we considered two dichotomisations and the ordinal specification. Under the first dichotomisation, denoted by $D_2$, the outcome was coded as 0 for very good or good health and 1 for fair, bad or very bad health. Under the second dichotomisation, denoted by $D_3$, the outcome was coded as 0 for very good, good or fair health and 1 for bad or very bad health. Thus, the two dichotomisations differ in the classification of the central ``fair'' category. In the ordinal specification, the original five response categories were retained. 

Age group and education level are the categorical stratification variables used in the analyses. They enter all models as fixed effects, with the youngest age group and the lowest education level as reference categories. For respondent $i$, let $a_i$ and $e_i$ denote the corresponding age group and education level, respectively. For each dichotomisation $D_c$, with $c \in \{2,3\}$, the binary model is specified as:
\begin{align*}
Y_i^{D_c} \mid \pi_{a_i e_i r_i}^{D_c}
&\sim \mathrm{Bernoulli}\left(\pi_{a_i e_i r_i}^{D_c}\right),\\
\mathrm{logit}\left(\pi_{a_i e_i r_i}^{D_c}\right)
&= \beta_0^{D_c} + \beta_{a_i}^{\mathrm{age}, D_c} + \beta_{e_i}^{\mathrm{edu}, D_c} + \theta_{r_i}^{\mathrm{nuts}, D_c},
\end{align*}
where $\theta_r^{\mathrm{nuts}, D_c}$ denotes the region-level spatial random effect for region $r$.

For the ordinal specification, the model is given by:
\begin{align*}
Y_i \mid \boldsymbol{\pi}_{a_i e_i r_i}
&\sim \mathrm{Categorical}\left(\boldsymbol{\pi}_{a_i e_i r_i}\right), \\
\mathrm{logit}\left(\gamma_{j a_i e_i r_i}\right)
&= \kappa_j - \beta_{a_i}^{\mathrm{age}} - \beta_{e_i}^{\mathrm{edu}} - \theta_{r_i}^{\mathrm{nuts}}.
\end{align*}
Here, $\gamma_{j a e r} = P(Y_i \leq j \mid a_i = a, e_i = e, r_i = r)$ for $j=1,\dots,J-1$. As described in Section~\ref{subsec:ordinal}, the category probabilities in $\boldsymbol{\pi}_{aer}$ are recovered from the cumulative probabilities, and the minus-sign convention implies that positive age, education and regional effects are associated with less favourable health.

All models are estimated in a Bayesian framework. Non-informative priors are assigned to the non-spatial model parameters, whereas the regional effects are assigned a Leroux conditional autoregressive prior to account for spatial dependence between neighbouring regions \citep{BeltranSanchez2024}. The models were implemented in \texttt{R} using the MCMC package \texttt{NIMBLE} \citep{Valpine2017}. All code used for the statistical analysis is available as a fully reproducible script in the \href{https://github.com/bsmiguelangel/the-risks-of-dichotomising-ordinal-outcomes}{project \texttt{GitHub} repository}.

\section{Results} \label{sec:results}

The results compare the ordinal model with the two dichotomised specifications for male respondents in Western Europe. We first examine descriptive regional patterns and model-based age, education and spatial effects, and then compare the poststratified regional estimates obtained from the models.

\subsection{Descriptive spatial patterns}

\begin{figure}[H]
    \vspace{-0.5cm}
    \hspace{-1.25cm}
    \includegraphics[width=16cm, trim={0cm 4cm 0cm 4cm}, clip]{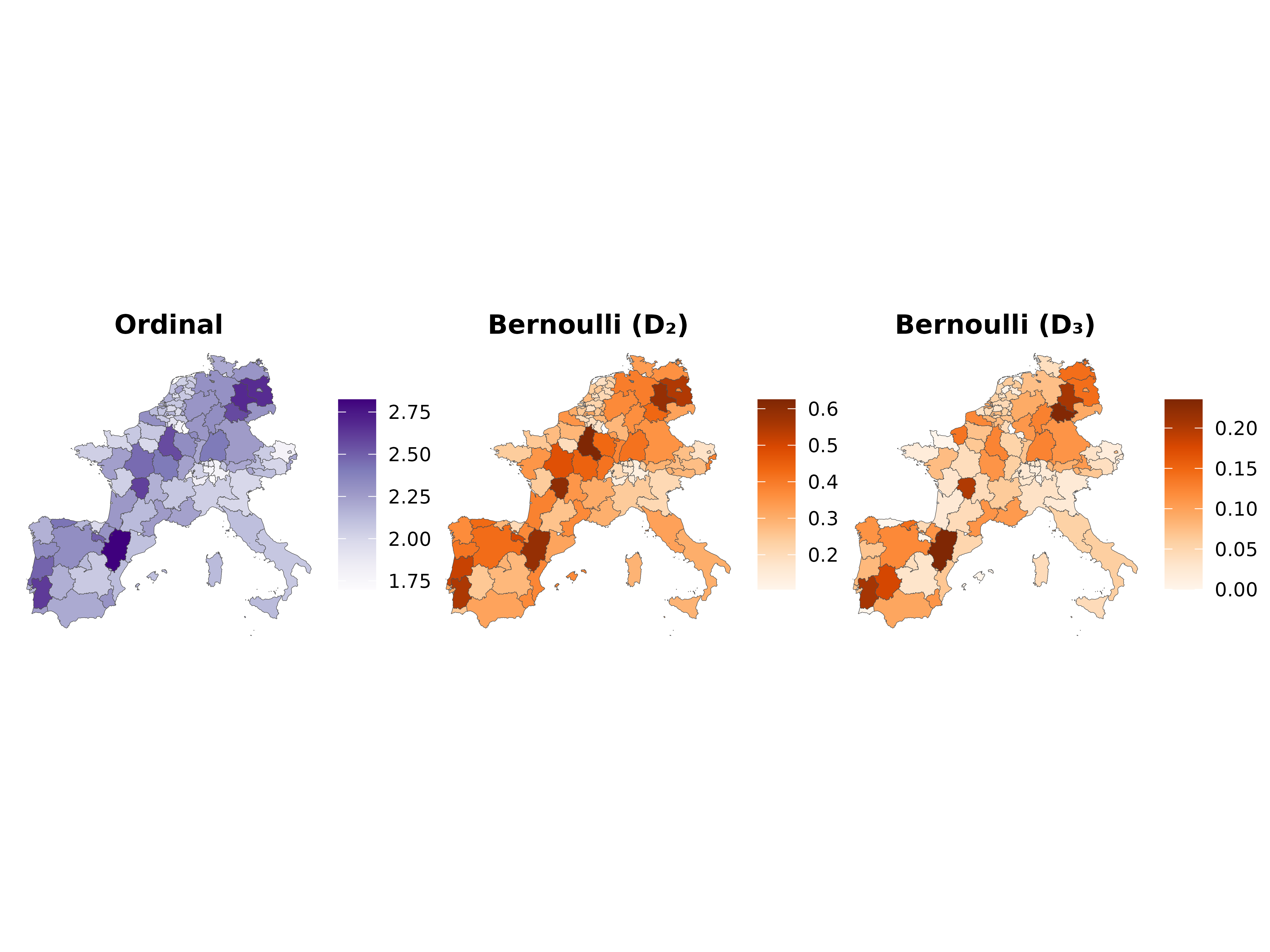}
    \vspace{-1.25cm}
    \caption{Unweighted raw regional averages of SRH among male respondents.}
    \label{fig:descriptive_man}
\end{figure}

Figure~\ref{fig:descriptive_man} shows unweighted raw regional averages of SRH, based directly on the observed survey responses before model-based smoothing or poststratification. They should therefore be interpreted as exploratory, since regional averages based only on survey respondents can be sensitive to small regional sample sizes and to differences between the observed sample composition and the target population. In the ordinal map, higher values indicate less favourable reported health on the original five-point scale. In the Bernoulli maps, regional averages correspond to the observed prevalence of the binary outcome defined by each dichotomisation.

The comparison shows that the spatial description of SRH depends on how the central ``fair'' category is treated: the regions with relatively less favourable SRH are not identical across the ordinal coding and the two dichotomisations. This descriptive comparison illustrates that the treatment of the central ``fair'' category affects the spatial pattern being described.

\subsection{Age and education effects}

\begin{figure}[H]
    \hspace{2cm}
    \includegraphics[width=10cm]{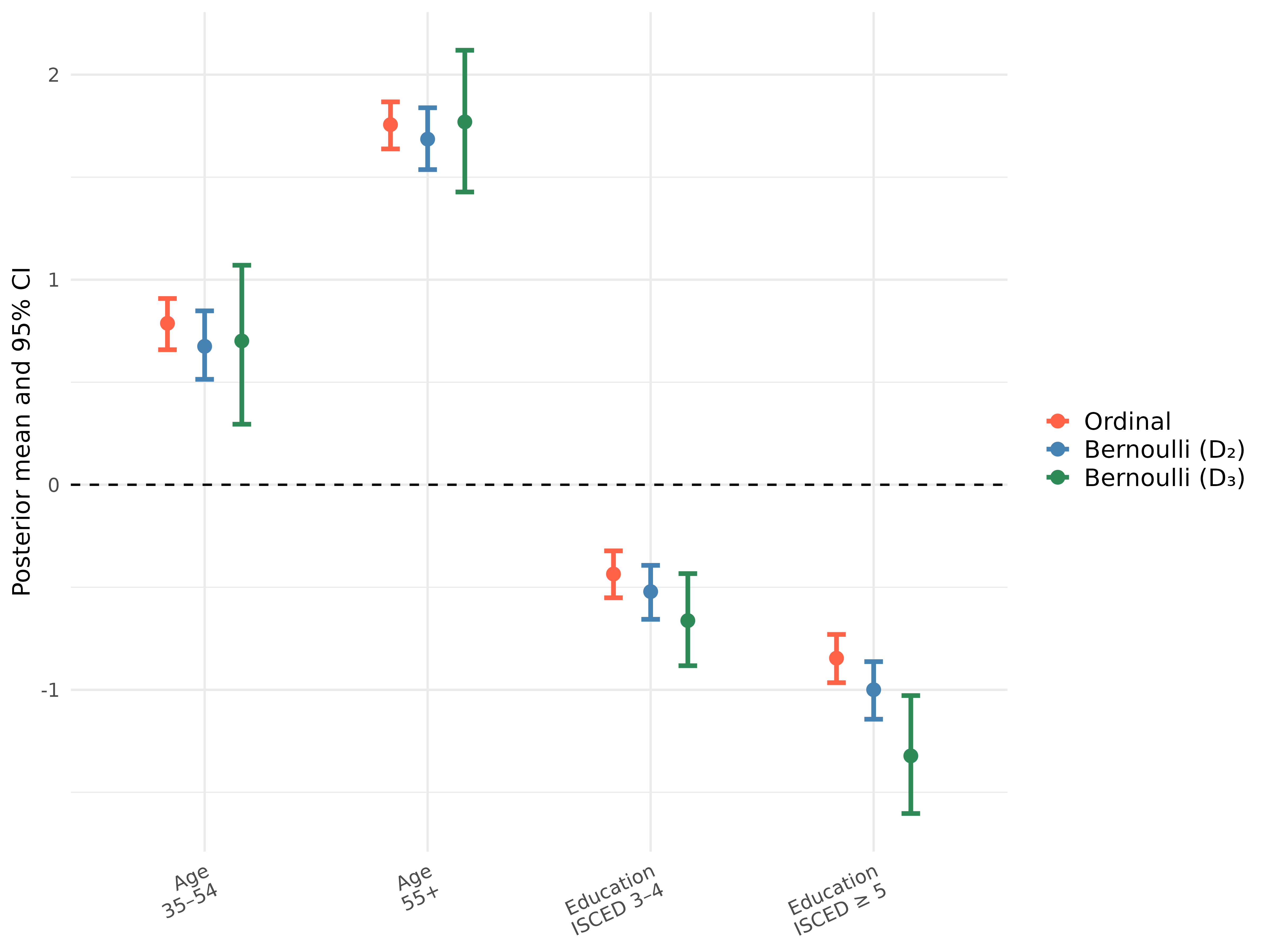}
    \caption{Estimated age and education effects among male respondents. Effects are relative to age 15--34 and ISCED $\leq 2$.}   
    \label{fig:fixed_man}
\end{figure}

Figure~\ref{fig:fixed_man} summarises the effects of the stratification variables in the three model specifications.

The direction of the effects is consistent across models. Since the ordinal and Bernoulli effects have the same direction of interpretation under the sign convention used here, as shown in Supplementary Material~A, positive effects correspond to less favourable SRH. The 95\% credible intervals exclude zero for all non-reference age and education categories. Older age is associated with less favourable SRH, especially among respondents aged 55 years or older. For education, the middle- and high-education groups are associated with better SRH than the low-education group, and this association is stronger for the highest education level. However, the magnitude and uncertainty of the effects differ across model specifications. In particular, the education gradient is more pronounced under $D_3$ than under $D_2$, reflecting the more restrictive definition of less favourable health in $D_3$. The ordinal model leads to the same qualitative conclusions while using the full five-category response scale.

\subsection{Spatial random effects}

\begin{figure}[H]
    \vspace{-0.5cm}
    \hspace{-1cm}
    \includegraphics[width=16cm]{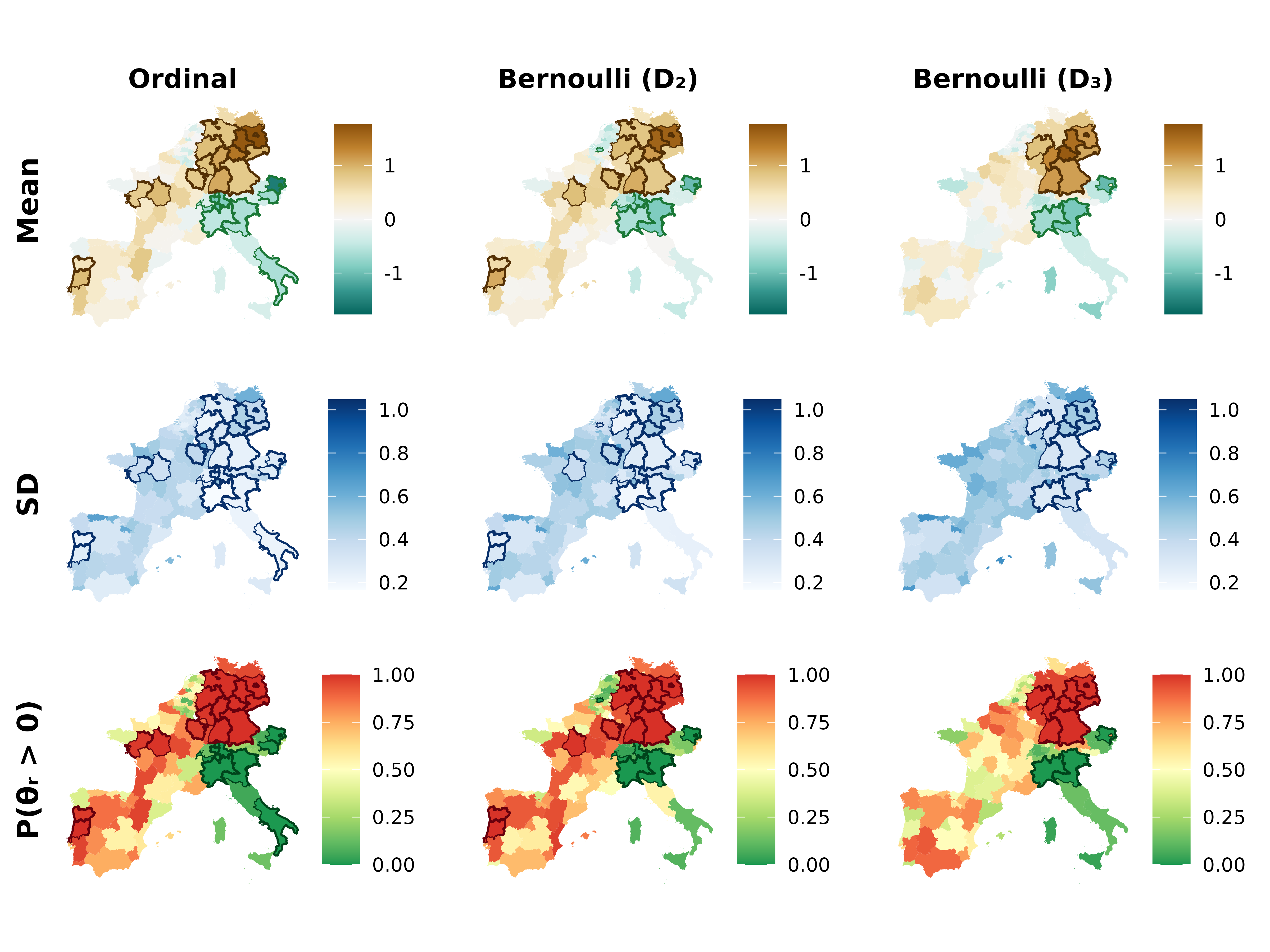}
    \vspace{-1cm}
    \caption{Posterior summaries of the region-level spatial random effects, $\theta_r$, among male respondents. Rows show the posterior mean, standard deviation and probability that the spatial effect is positive. Outlined regions have 95\% credible intervals excluding zero.}
    \label{fig:spatial_effect_man}
\end{figure}

Figure~\ref{fig:spatial_effect_man} summarises the posterior mean, standard deviation and probability of being positive for the region-level spatial random effects, $\theta_r$, under the three models. These effects capture residual geographical variation in SRH after accounting for age and education. Positive spatial effects indicate worse SRH and are shown in brown, whereas negative effects indicate better SRH and are shown in green. The probability maps show the posterior probability that each spatial effect is positive; values close to one indicate strong evidence of worse SRH, while values close to zero indicate strong evidence of better SRH. Outlined regions have 95\% credible intervals excluding zero.

The ordinal model reveals several regional differences in the spatial effects. For example, German regions have positive spatial effects, indicating worse SRH after accounting for age and education, whereas Italy, Austria and Switzerland tend to show negative spatial effects, indicating better SRH. These country-level contrasts are also reflected in the probability maps: several of these regions have 95\% credible intervals that exclude zero, supporting geographical differences in SRH across Western Europe.

The comparison with the Bernoulli models shows that these geographical conclusions depend on the dichotomisation used. The $D_2$ model remains closer to the ordinal model than the $D_3$ model. However, some regions with clear evidence of geographical differences under the ordinal model no longer have credible intervals excluding zero after dichotomisation, and the magnitude of the spatial effects also changes. This loss of evidence is more pronounced under $D_3$, where fewer regions have credible intervals excluding zero. Overall, the spatial effects show that both the geographical pattern and the evidence for regional differences are sensitive to the dichotomisation used.

\subsection{Poststratified regional estimates}

\begin{figure}[H]
    \vspace{-0.75cm}
    \hspace{-1.5cm}
    \includegraphics[width=16.5cm]{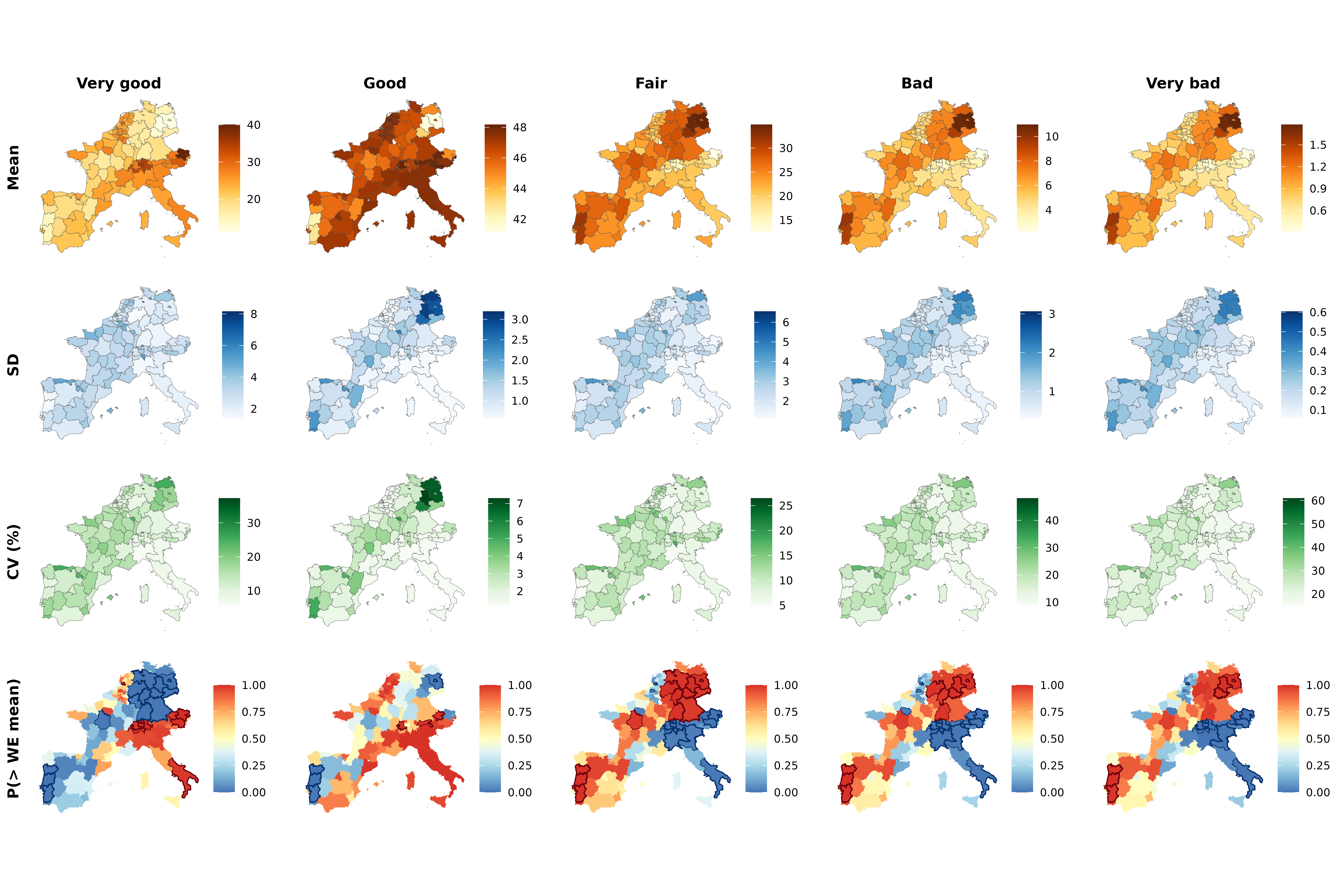}
    \vspace{-1.5cm}
    \caption{Poststratified regional percentages in each SRH category among male respondents under the ordinal model. Rows show the posterior mean, standard deviation, coefficient of variation and probability of exceeding the corresponding Western European mean. Outlined regions have 95\% prediction intervals excluding that mean.}
    \label{fig:prevalence_ordinal_man}
\end{figure}

Figure~\ref{fig:prevalence_ordinal_man} presents the poststratified regional percentages obtained from the ordinal model for each of the five SRH categories. This is one of the main advantages of retaining the ordinal scale: the model provides a separate population percentage for each response category, rather than a single binary prevalence. Unlike the spatial random effects, which describe residual regional variation after accounting for age and education, these poststratified estimates combine the modelled SRH probabilities for each age-by-education group with the population composition of each NUTS region to estimate the percentage of the regional population in each SRH category. 

The posterior mean percentages show that most of the population is distributed across the very good, good and fair categories, whereas bad and especially very bad health account for much smaller percentages. The uncertainty patterns differ across categories: posterior standard deviations are not determined only by the size of the estimated percentages, since the category probabilities are estimated jointly as part of the full ordinal response distribution. The coefficients of variation, which express uncertainty relative to the estimated percentage, are higher for the rarer categories, especially bad and very bad health. The exceedance-probability maps show which regions are likely to be above the Western European mean for each category.

\begin{figure}[H]
    \vspace{-0.75cm}
    \hspace{-1.75cm}
    \includegraphics[width=17cm]{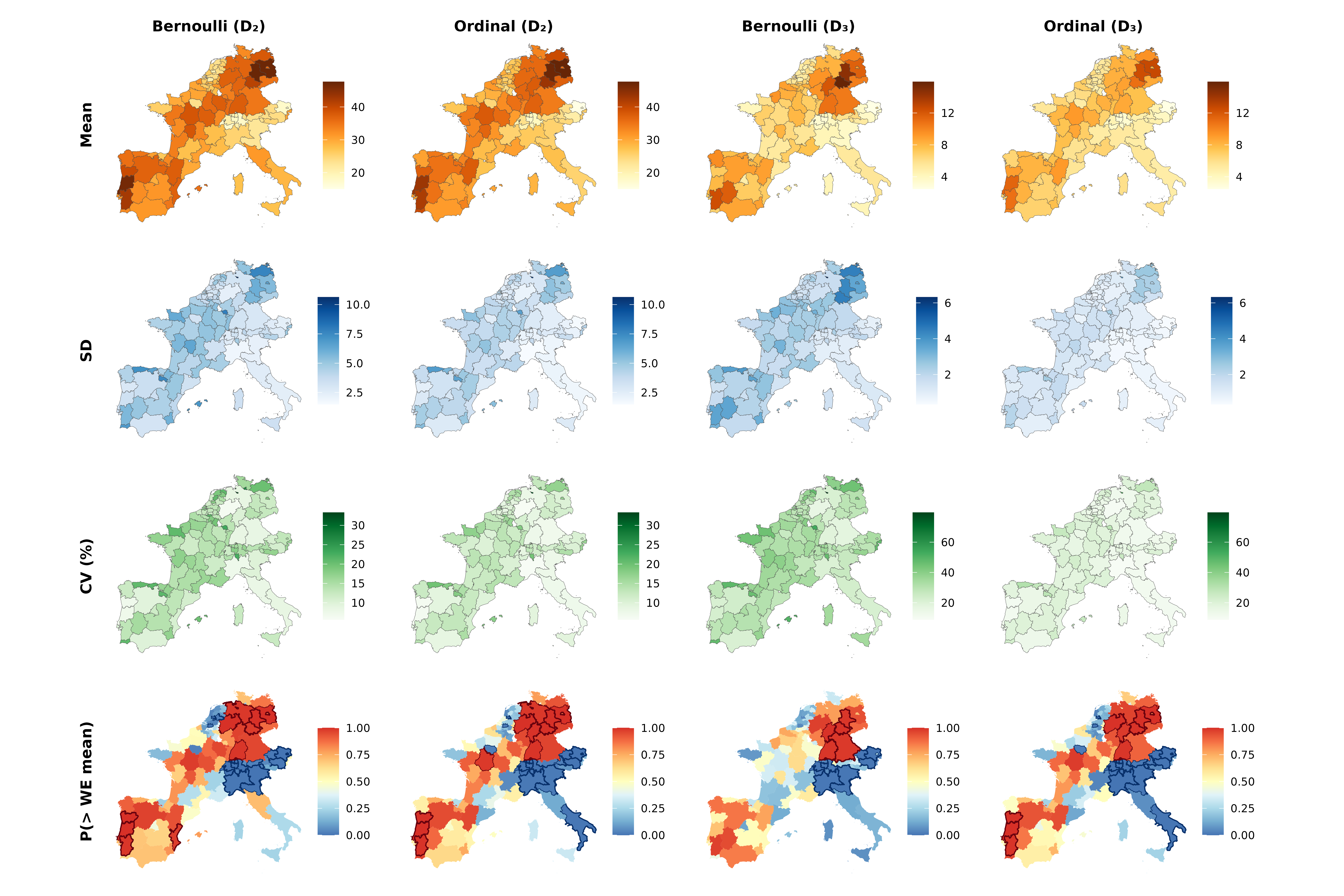}
    \vspace{-0.5cm}
    \caption{Comparison of poststratified regional prevalences from the Bernoulli models and the corresponding binary summaries derived from the ordinal model among male respondents. Outlined regions have 95\% prediction intervals excluding the corresponding Western European mean.}
    \label{fig:comparison_man}
\end{figure}

Figure~\ref{fig:comparison_man} compares the poststratified regional prevalences obtained directly from the Bernoulli models with the corresponding binary summaries derived from the ordinal model. In the ordinal model, these summaries are obtained by adding the poststratified percentages of the relevant SRH categories within each region.

The two dichotomisations lead to different prevalence levels, uncertainty patterns and exceedance maps. Posterior mean prevalences are higher under $D_2$ than under $D_3$, but the more relevant difference lies in which regions are identified as above the Western European mean. For example, several Portuguese regions and the Valencian Community are outlined as above the Western European mean under $D_2$ but not under $D_3$, while some German regions change their exceedance status between dichotomisations. These changes show that the regional classification depends on whether the central ``fair'' category is included in the binary outcome.

Within each dichotomisation, the posterior mean prevalences from the Bernoulli model and the corresponding ordinal aggregation are broadly similar. This shows that the ordinal model can recover the main binary prevalence patterns when its category-specific estimates are combined according to the same cut-off point. However, the uncertainty summaries differ: the ordinal aggregations generally show smaller posterior standard deviations and coefficients of variation than the corresponding Bernoulli estimates. The exceedance-probability maps are also not identical, indicating that similar posterior means do not necessarily imply the same regional evidence for exceeding the Western European mean. A direct comparison of the differences between the Bernoulli prevalences and the corresponding ordinal-model aggregations is reported in Supplementary Material~B. The posterior mean differences are generally small, with 95\% prediction intervals including zero in almost all regions. This supports the conclusion that the ordinal model can closely recover the binary prevalences when required, while retaining the full five-category SRH distribution.

\begin{figure}[H]
    \hspace{-0.5cm}
    \includegraphics[width=14cm]{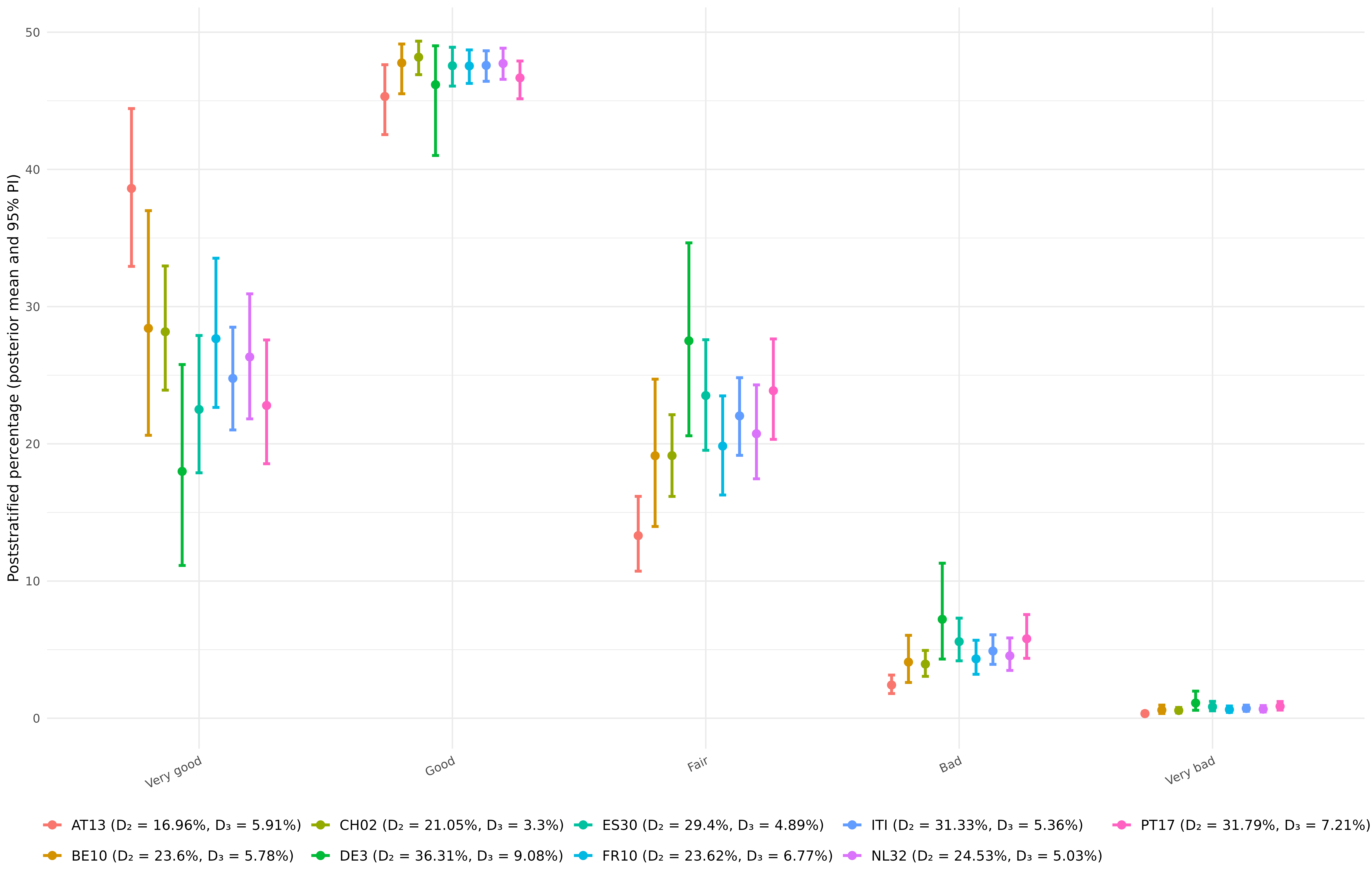}
    \caption{Poststratified ordinal profiles for selected NUTS regions among male respondents. The legend reports posterior means of the poststratified population prevalences for $D_2$ and $D_3$.}
    \label{fig:ordinal_profiles_man}
\end{figure}

Finally, Figure~\ref{fig:ordinal_profiles_man} displays the full poststratified ordinal profiles for the NUTS regions containing the capital city of each country. The selected regions are AT13 (Vienna), BE10 (Brussels-Capital Region), CH02 (Espace Mittelland), DE3 (Berlin), ES30 (Community of Madrid), FR10 (Ile-de-France), ITI (Central Italy), NL32 (North Holland) and PT17 (Lisbon Metropolitan Area). Each profile shows the estimated percentage of the regional male population in each of the five SRH categories, with 95\% prediction intervals.

A binary prevalence summarises only the combined percentage above a chosen cut-off, whereas the ordinal profile shows how that percentage is distributed across the original response categories. Thus, binary prevalences can hide important differences in the underlying SRH distribution: similar binary prevalences may correspond to different ordinal profiles, and differences in binary prevalences can be interpreted more clearly by identifying which SRH categories account for them. For example, AT13 (Vienna) and DE3 (Berlin) have broadly similar estimates for good health, with overlapping prediction intervals, but they differ more clearly in other categories: Vienna has a higher percentage of very good health, whereas Berlin has higher percentages of fair and bad health. This final comparison illustrates the additional information provided by the ordinal model: it provides the binary prevalences when required, while retaining the full five-category distribution for substantive interpretation.

\section{Discussion} \label{sec:discussion}

This paper examined the consequences of dichotomising an ordinal health outcome in a spatial analysis of SRH in Western Europe. Using male respondents from ESS11, we compared an ordinal cumulative logit model with two Bernoulli models based on alternative cut-offs that differ only in the treatment of the central ``fair'' category. Older age and lower education were consistently associated with less favourable SRH across all specifications. However, the magnitude and uncertainty of the estimated effects, the residual spatial patterns and the poststratified population estimates depended on how the ordinal outcome was treated. 

The main finding is that dichotomisation is not a simple recoding. By assigning the central ``fair'' category to one side or the other of the binary outcome, the two dichotomisations change the scale and interpretation of less favourable health. This affects the geographical patterns obtained from the models. In particular, the spatial random effects show that the selected cut-off can alter the posterior evidence for residual regional differences: some regions with clear evidence of better or worse SRH under the ordinal model no longer show the same evidence after dichotomisation, especially under the more restrictive $D_3$ definition. 

The poststratified results provide a complementary perspective. The ordinal model yields population percentages for each of the five original SRH categories, whereas the Bernoulli models yield prevalences for the chosen binary outcomes. When the relevant category-specific percentages from the ordinal model are combined according to $D_2$ or $D_3$, the resulting prevalences closely reproduce those obtained directly from the Bernoulli models. The difference maps show little region-level evidence of discrepancies between these two approaches after aggregation. This shows that aggregating the ordinal estimates gives results that are consistent with the direct binary models, while retaining the category-specific information discarded by those models.

These findings have broader implications for health-inequality research beyond this specific case study. Previous work has examined whether analyses of SRH are sensitive to treating the response as dichotomous or categorical \citep{Manor2000}. However, dichotomised SRH indicators are not merely a matter of methodological convenience. They are widely used in social epidemiology and health-inequality research, including comparative monitoring based on ESS data \citep{EuroHealthNetCHAIN2025}. Prevalence measures such as the percentage of the population reporting less than good health provide a substantively meaningful way of summarising population health. This issue also relates to broader work on the measurement of health inequalities. \citet{Eikemo2009} showed that variation in relative health inequalities may partly depend on the underlying prevalence of the health outcome, and argued that relative measures should be interpreted together with absolute levels and absolute inequalities. Although our focus is different, the common concern is whether apparent differences in health inequalities between populations reflect substantive social differences or are partly produced by properties of the measures and modelling decisions used to describe them.

Our results extend this concern to a spatial setting. The way health is measured is not a purely technical step that occurs after the substantive research question has been formulated: it can influence the social and geographical patterns observed in the analysis. In this study, changing the treatment of the central ``fair'' category changes the scale of less favourable health and affects which regions are identified as relatively disadvantaged. This has direct implications for comparative health-inequality research and, potentially, for policy prioritisation, as regional classifications may depend on how the outcome is modelled. The consequences are likely to be most pronounced when the central category contains a substantial proportion of respondents, as is the case for ``fair'' health in many survey settings.

These considerations are also relevant for policy communication. Dichotomisation remains attractive because binary prevalence measures are readily understandable to policymakers and non-specialist audiences. The implication is not that binary indicators should no longer be reported. Rather, when feasible, the ordinal information should be preserved in the underlying analysis, since familiar binary prevalence estimates can subsequently be obtained by aggregating categories. This combines statistically more complete modelling with simple, interpretable indicators for monitoring and policy communication. In this context, category-specific maps and regional profiles can show whether less favourable health is mainly driven by the central ``fair'' category, by the more severe ``bad'' and ``very bad'' categories, or by differences across several response categories. 

Taken together, these results help answer a practical question for applied health-inequalities research: how much confidence should be placed in conclusions based on dichotomised SRH? Our findings do not suggest that this established research tradition should be dismissed. Instead, they suggest a more specific interpretation: the fundamental age and educational gradients appear relatively robust across specifications, whereas their magnitude and uncertainty, and some of the geographical conclusions, are sensitive to how the outcome is modelled.

The main limitation of this study is that the complex ESS sampling design is not fully accounted for. Sampling procedures vary across countries, and detailed information on the exact sampling designs is difficult to obtain consistently across all countries. Our model-based analysis adjusts for the most common stratification variables, which are age, education and region \citep{ESSWeighting}, by including them in the model and through poststratification, but it does not explicitly incorporate all country-specific design features. As a result, the empirical regional estimates and their uncertainty should be interpreted with some caution. However, the central methodological comparison remains informative because the ordinal and dichotomised models are fitted under the same assumptions; the observed differences between them therefore reflect the treatment of the ordinal outcome. The models also include main effects for age group, education level and region, but do not include age-by-education interactions. Such interactions may be relevant for understanding how educational inequalities in SRH vary across age groups. We deliberately use this main-effects specification to keep the modelling framework simple and to focus on the consequences of dichotomising the outcome. Finally, the analysis is based on one ESS round and a selected set of Western European countries, so the empirical patterns should not be interpreted as universal features of SRH in Europe.

In conclusion, this analysis shows that dichotomising an ordinal outcome is a consequential modelling decision, particularly in spatial settings. Collapsing ordered categories may simplify the analysis, but it can also change residual spatial patterns, uncertainty and the evidence for regional differences. Binary indicators nevertheless remain useful for monitoring and policy communication, but the selected cut-off should be reported and justified, and sensitivity to alternative cut-offs should be assessed. When feasible, ordinal models can provide a useful complementary approach because they retain information across all outcome categories, avoid the need to select a single binary cut-off, and can still produce binary summaries when required.
\vspace*{1pc}

\noindent {\bf{Acknowledgements}}

\noindent The authors acknowledge the funding of Ministerio de Ciencia, Innovación y Universidades of Spain (MCIN/AEI/10.13039/501100011033/ FEDER, UE) and the European Regional Development Fund. 
\vspace*{1pc}

\noindent {\bf{Funding}}

\noindent This paper is part of the project PID2022-136455NB-I00, funded by Ministerio de Ciencia, Innovación y Universidades of Spain (MCIN/AEI/10.13039/50\\ 1100011033/ FEDER, UE) and the European Regional Development Fund. 
\vspace*{1pc}

\noindent {\bf{Declaration of competing interest}}

\noindent The authors declare no potential conflicts of interest with respect to the research, authorship, and/or publication of this article.
\vspace*{1pc}

\noindent {\bf{Data availability statement}}

\noindent The ESS data used in this article are publicly available through the ESS Data Portal. The population data used for poststratification are publicly available from Eurostat Census 2021 tables. All code used for the statistical analysis is available as a fully reproducible script in the \href{https://github.com/bsmiguelangel/the-risks-of-dichotomising-ordinal-outcomes}{project \texttt{GitHub} repository}.
\vspace*{1pc}

\noindent {\bf{ORCID iDs}}

\noindent 
Beltrán-Sánchez, MA \orcidlink{0000-0001-9450-2973} \url{https://orcid.org/0000-0001-9450-2973} \\ 
Martínez-Beneito, MA \orcidlink{0000-0001-8406-8050} \url{https://orcid.org/0000-0001-8406-8050} \\
Eikemo, Terje \orcidlink{0000-0001-6809-7011} \url{https://orcid.org/0000-0001-6809-7011} \\
Martino, Sara \orcidlink{0000-0003-4326-9029} \url{https://orcid.org/0000-0003-4326-9029} \\
Riebler, Andrea \orcidlink{0009-0004-3998-4032} \url{https://orcid.org/0009-0004-3998-4032}
\vspace*{1pc}

\noindent {\bf{Supplementary material}}

\noindent The Supplementary Material for this article is available online. It contains the connection between the ordinal and binary models and additional results for male and female respondents.







\newpage

\setcounter{figure}{0}
\renewcommand{\thefigure}{S\arabic{figure}}

\setcounter{table}{0}
\renewcommand{\thetable}{S\arabic{table}}


\phantomsection
\addcontentsline{toc}{section}{Supplementary Material}
\section*{Supplementary Material}

\phantomsection
\addcontentsline{toc}{subsection}{A. Connection between the ordinal and binary models}
\subsection*{A. Connection between the ordinal and binary models}
\label{supp:ordinal_binary_or}

This supplementary section clarifies how the effects in the ordinal cumulative logit model can be interpreted in relation to those in a Bernoulli logistic regression model obtained after dichotomising the ordinal outcome. Categories are ordered from better to worse health, with $Y_i \in \{1,\dots,J\}$. For a cut-off point $c$, the binary outcome is defined as:
\begin{equation*}
Y_i^{D_c} =
\begin{cases}
0, & Y_i \leq c,\\
1, & Y_i > c,
\end{cases}
\end{equation*}
so that $Y_i^{D_c} = 1$ denotes membership of the less favourable health group. 

In the ordinal model in the main text, the cumulative probability up to category $c$ for stratum $z$ and region $r$ is $\gamma_{czr} = P(Y_i \leq c \mid z_i = z, r_i = r) = \sum_{k \leq c} \pi_{k z r}$. In the Bernoulli model in the main text, the modelled probability is $\pi_{zr}^{D_c} = P(Y_i^{D_c}=1 \mid z_i = z, r_i = r)$; that is, the probability of being above the cut-off point. The corresponding probability obtained from the ordinal model is $1 - \gamma_{czr}$.

Under the ordinal cumulative logit model, $\mathrm{logit}\left(\gamma_{czr}\right) = \kappa_c - \beta_z - \theta_r$. Since the dichotomised outcome refers to the complementary event $Y_i > c$, the logit of the corresponding probability derived from the ordinal model is $\mathrm{logit}\left(1 - \gamma_{czr}\right) = -\kappa_c + \beta_z + \theta_r$. Therefore, for a given cut-off point $c$, the expression derived from the ordinal model has the same structure as a Bernoulli logistic regression model for the dichotomised outcome. In this sense, the ordinal model induces a binary formulation in which the intercept corresponds to $-\kappa_c$, while the stratum and regional effects have the same direction of interpretation as in the Bernoulli model. Positive values of $\beta_z$ or $\theta_r$ are associated with higher probabilities of being above the cut-off point, and therefore with less favourable health, as in the Bernoulli model.

The same relationship can be expressed in terms of odds ratios. First consider the Bernoulli model fitted to the dichotomised outcome. For a given cut-off point $c$, the stratum effect compares stratum $z$ with the reference stratum within the same region $r$:
\begin{equation*}
\beta_z^{D_c} = \mathrm{logit}\left(\pi_{zr}^{D_c}\right) - \mathrm{logit}\left(\pi_{1r}^{D_c}\right).
\end{equation*}
Therefore, $\exp(\beta_z^{D_c})$ is the odds ratio comparing stratum $z$ with the reference stratum for belonging to the less favourable health group:
\begin{equation*}
\exp(\beta_z^{D_c}) = \frac{\pi_{zr}^{D_c}/\left(1-\pi_{zr}^{D_c}\right)}
{\pi_{1r}^{D_c}/\left(1-\pi_{1r}^{D_c}\right)}.
\end{equation*}
In the ordinal model, the cumulative logit specification gives:
\begin{equation*}
\beta_z = \mathrm{logit}\left(\gamma_{c1r}\right)
- \mathrm{logit}\left(\gamma_{czr}\right) = \mathrm{logit}\left(1-\gamma_{czr}\right)
- \mathrm{logit}\left(1-\gamma_{c1r}\right).
\end{equation*}
Thus, for a given cut-off point $c$, the ordinal effect $\beta_z$ has the same odds-ratio interpretation as the stratum effect in the corresponding Bernoulli model:
\begin{equation*}
\exp(\beta_z) = \frac{(1-\gamma_{czr})/\gamma_{czr}}
{(1-\gamma_{c1r})/\gamma_{c1r}}.
\end{equation*}
The same reasoning applies to the regional effects. In the Bernoulli model, the regional effect compares regions $r$ and $s$ within the same stratum $z$:
\begin{equation*}
\theta_r^{D_c} - \theta_s^{D_c} = \mathrm{logit}\left(\pi_{zr}^{D_c}\right) - \mathrm{logit}\left(\pi_{zs}^{D_c}\right).
\end{equation*}
Therefore, $\exp(\theta_r^{D_c} - \theta_s^{D_c})$ is the odds ratio comparing regions $r$ and $s$ for belonging to the less favourable health group:
\begin{equation*}
\exp(\theta_r^{D_c} - \theta_s^{D_c}) =
\frac{\pi_{zr}^{D_c}/\left(1-\pi_{zr}^{D_c}\right)}
{\pi_{zs}^{D_c}/\left(1-\pi_{zs}^{D_c}\right)}.
\end{equation*}
In the ordinal model, this comparison is expressed as:
\begin{equation*}
\theta_r-\theta_s = \mathrm{logit}\left(\gamma_{czs}\right) - \mathrm{logit}\left(\gamma_{czr}\right) = \mathrm{logit}\left(1-\gamma_{czr}\right) - \mathrm{logit}\left(1-\gamma_{czs}\right).
\end{equation*}
Thus, $\exp(\theta_r - \theta_s)$ has the same odds-ratio interpretation as the corresponding regional effect in the Bernoulli model:
\begin{equation*}
\exp(\theta_r - \theta_s) = \frac{(1-\gamma_{czr})/\gamma_{czr}}
{(1-\gamma_{czs})/\gamma_{czs}}.
\end{equation*}
These relationships explain the role of the minus sign in the ordinal cumulative logit predictor. Because the dichotomised outcome is defined as being above the cut-off point, positive ordinal effects correspond to higher odds of belonging to the less favourable health group. The stratum and regional effects in the ordinal model are therefore connected in interpretation to those in the Bernoulli model for the same cut-off point. This connection concerns the interpretation of the effects. It does not imply that the ordinal model and the Bernoulli model fitted separately to the data must produce identical estimates. The Bernoulli model uses only the dichotomised outcome, whereas the ordinal model uses the full ordered response distribution.

\phantomsection
\addcontentsline{toc}{subsection}{B. Additional results}
\subsection*{B. Additional results}
\label{supp:additional_results}

\begin{figure}[H]
    \hspace{-0.5cm}
    \includegraphics[width=14.5cm, trim={0cm 1cm 0cm 1cm}, clip]{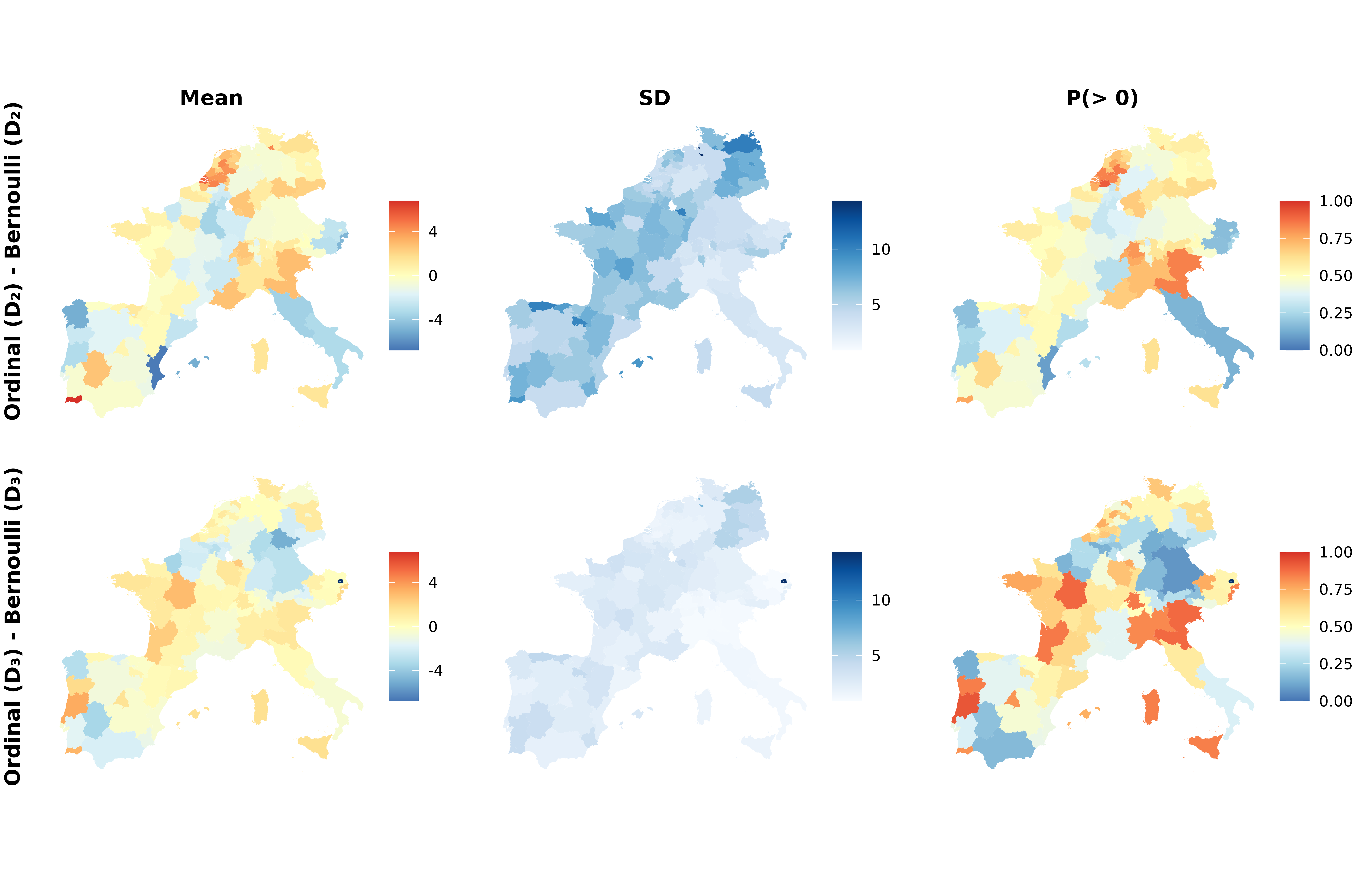}
    \vspace{-0.3cm}
    \caption{Differences between ordinal-model aggregations and Bernoulli poststratified regional prevalences among male respondents. Columns show the posterior mean difference, standard deviation and the probability that the difference is positive. Positive mean values indicate larger ordinal-model estimates. Outlined regions indicate 95\% prediction intervals for the difference excluding zero.}
    \label{fig:differences_man}
\end{figure}

\begin{figure}[H]
    \vspace{-0.5cm}
    \hspace{-1.25cm}
    \includegraphics[width=16cm, trim={0cm 4cm 0cm 4cm}, clip]{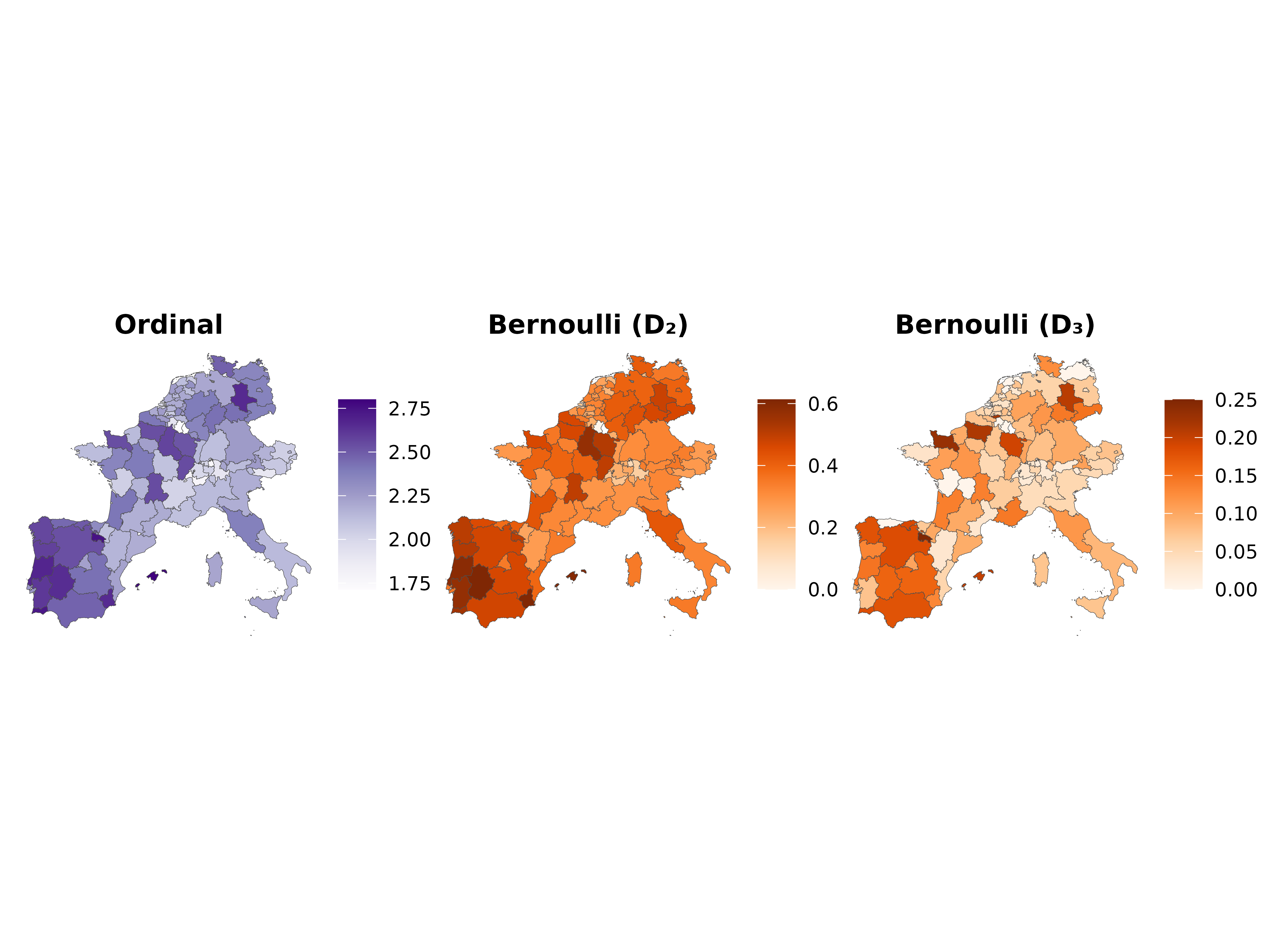}
    \vspace{-1.25cm}
    \caption{Unweighted raw regional averages of SRH among female respondents.}
    \label{fig:descriptive_woman}
\end{figure}

\begin{figure}[H]
    \hspace{2cm}
    \includegraphics[width=10cm]{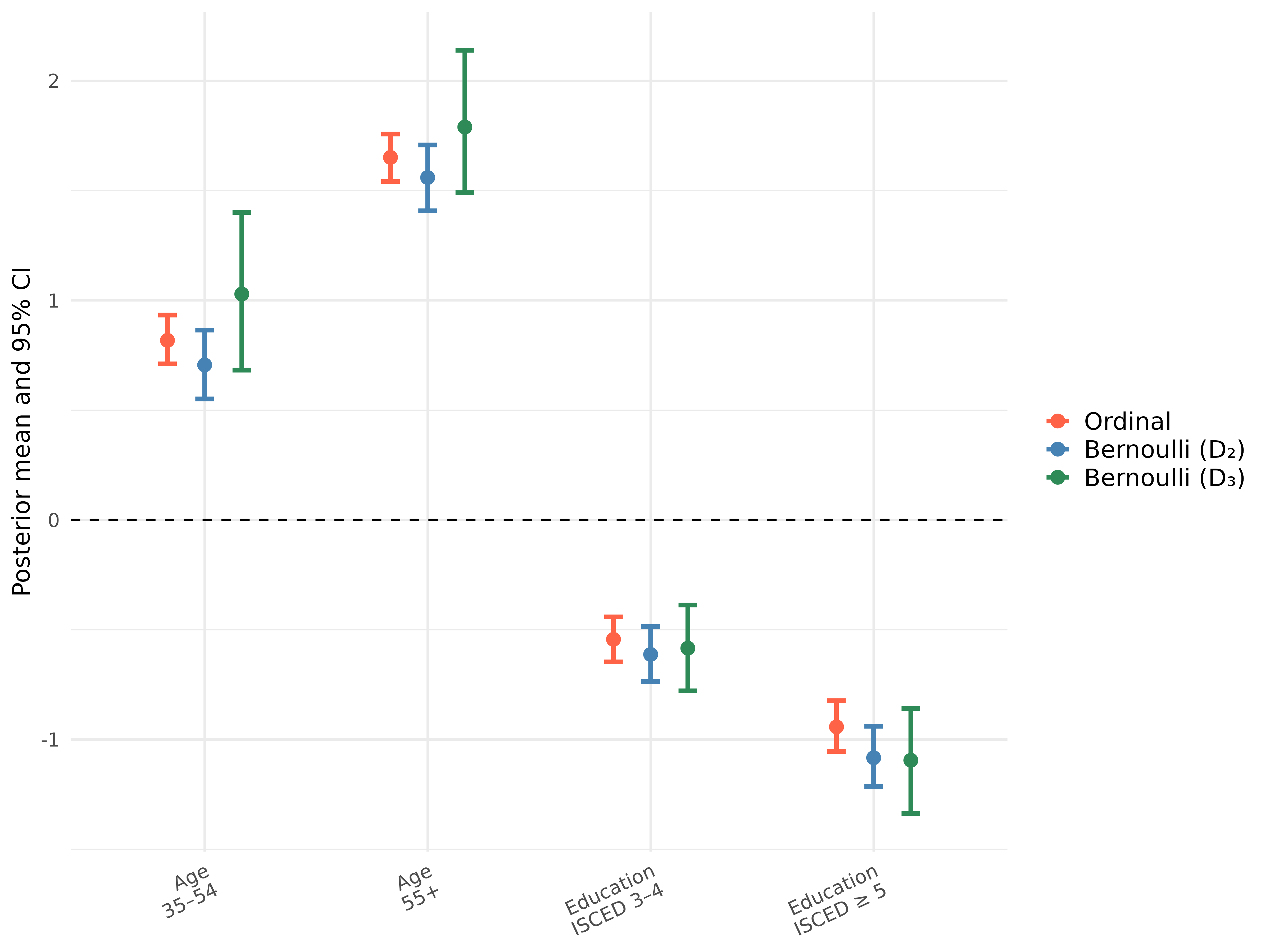}
    \caption{Estimated age and education effects among female respondents. Effects are relative to age 15--34 and ISCED $\leq 2$.}
    \label{fig:fixed_woman}
\end{figure}

\vspace{-0.5cm}

\begin{figure}[H]
    \vspace{-0.5cm}
    \hspace{-1cm}
    \includegraphics[width=16cm]{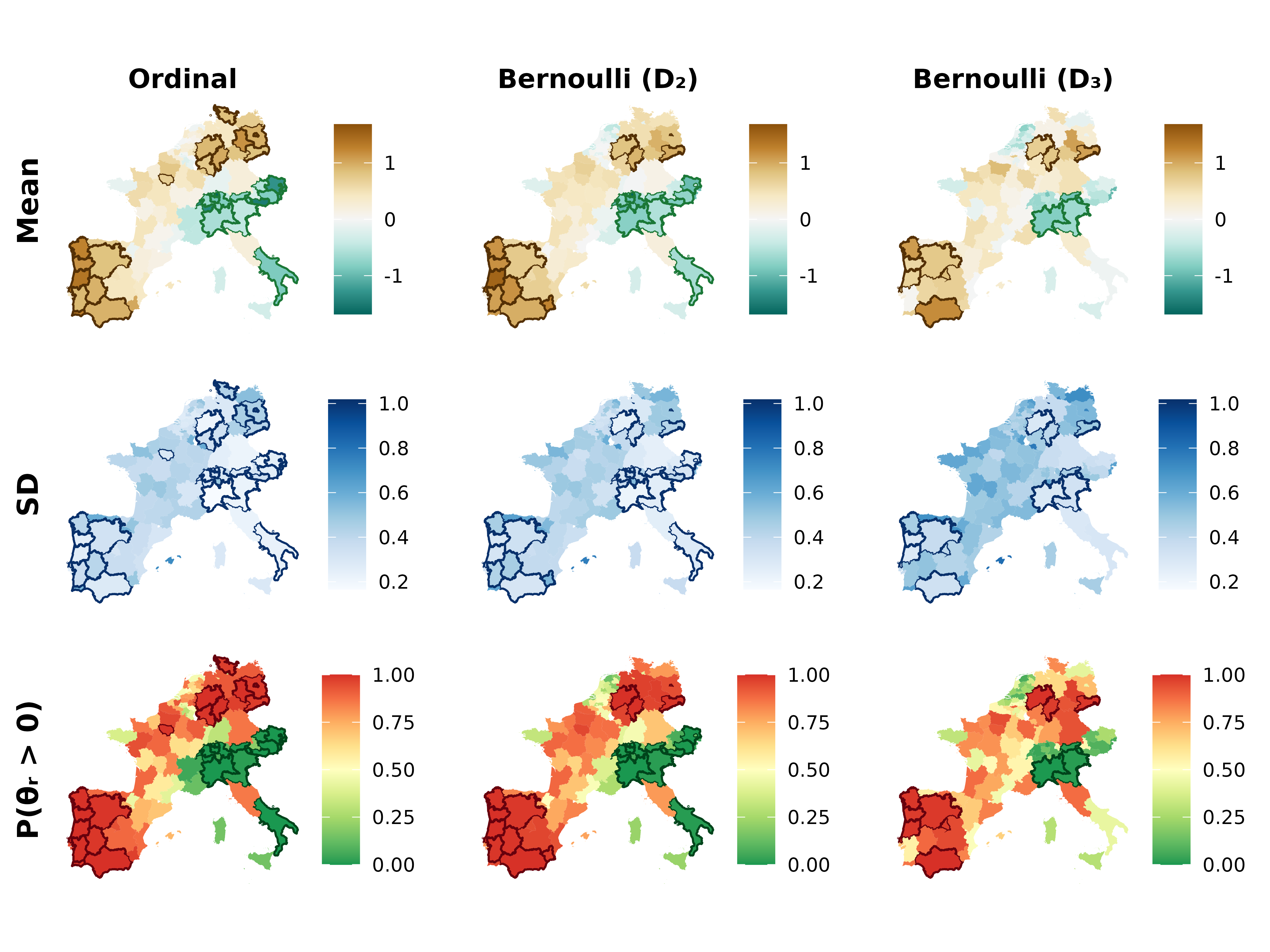}
    \vspace{-1cm}
    \caption{Posterior summaries of the region-level spatial random effects, $\theta_r$, among female respondents. Rows show the posterior mean, standard deviation and probability that the spatial effect is positive. Outlined regions have 95\% credible intervals excluding zero.}
    \label{fig:spatial_effect_woman}
\end{figure}

\begin{figure}[H]
    \vspace{-0.75cm}
    \hspace{-1.5cm}
    \includegraphics[width=16.5cm]{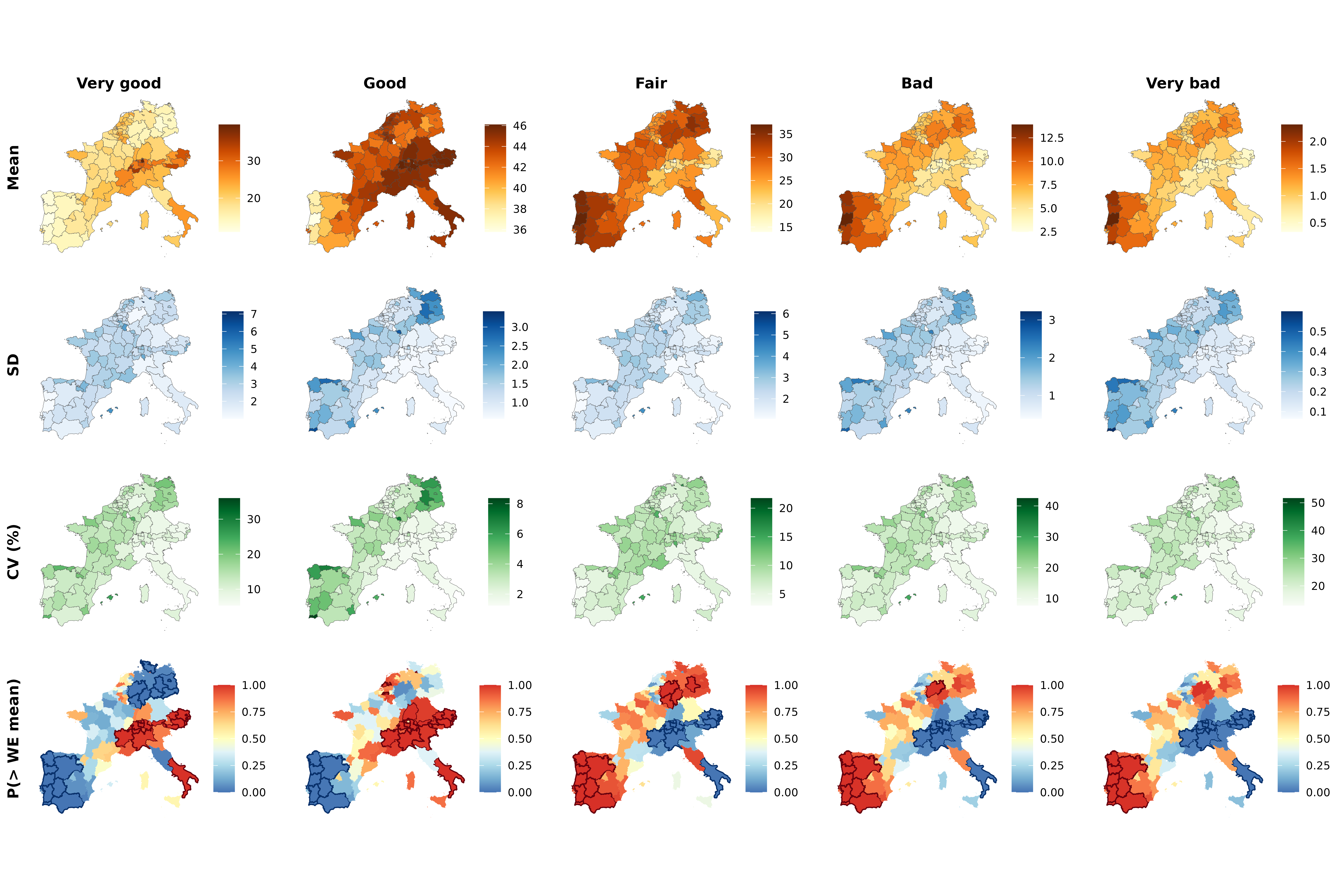}
    \vspace{-1.5cm}
    \caption{Poststratified regional percentages in each SRH category among female respondents under the ordinal model. Rows show the posterior mean, standard deviation, coefficient of variation and probability of exceeding the corresponding Western European mean. Outlined regions have 95\% prediction intervals excluding that mean.}
    \label{fig:prevalence_ordinal_woman}
\end{figure}

\begin{figure}[H]
    \vspace{-0.75cm}
    \hspace{-1.75cm}
    \includegraphics[width=17cm]{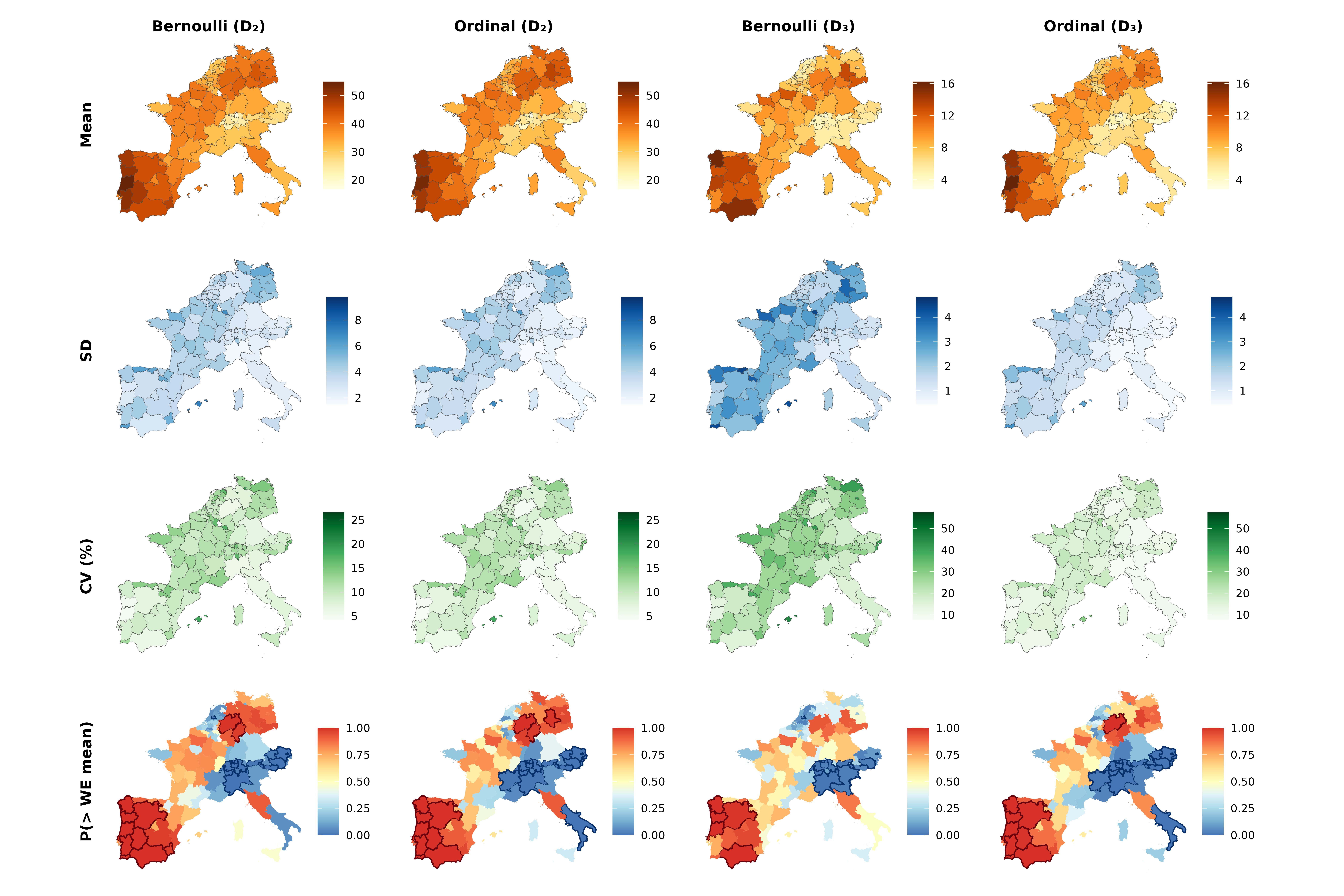}
    \vspace{-0.5cm}
    \caption{Comparison of poststratified regional prevalences from the Bernoulli models and the corresponding binary summaries derived from the ordinal model among female respondents. Outlined regions have 95\% prediction intervals excluding the corresponding Western European mean.}
    \label{fig:comparison_woman}
\end{figure}

\begin{figure}[H]
    \hspace{-0.5cm}
    \includegraphics[width=14.5cm, trim={0cm 1cm 0cm 1cm}, clip]{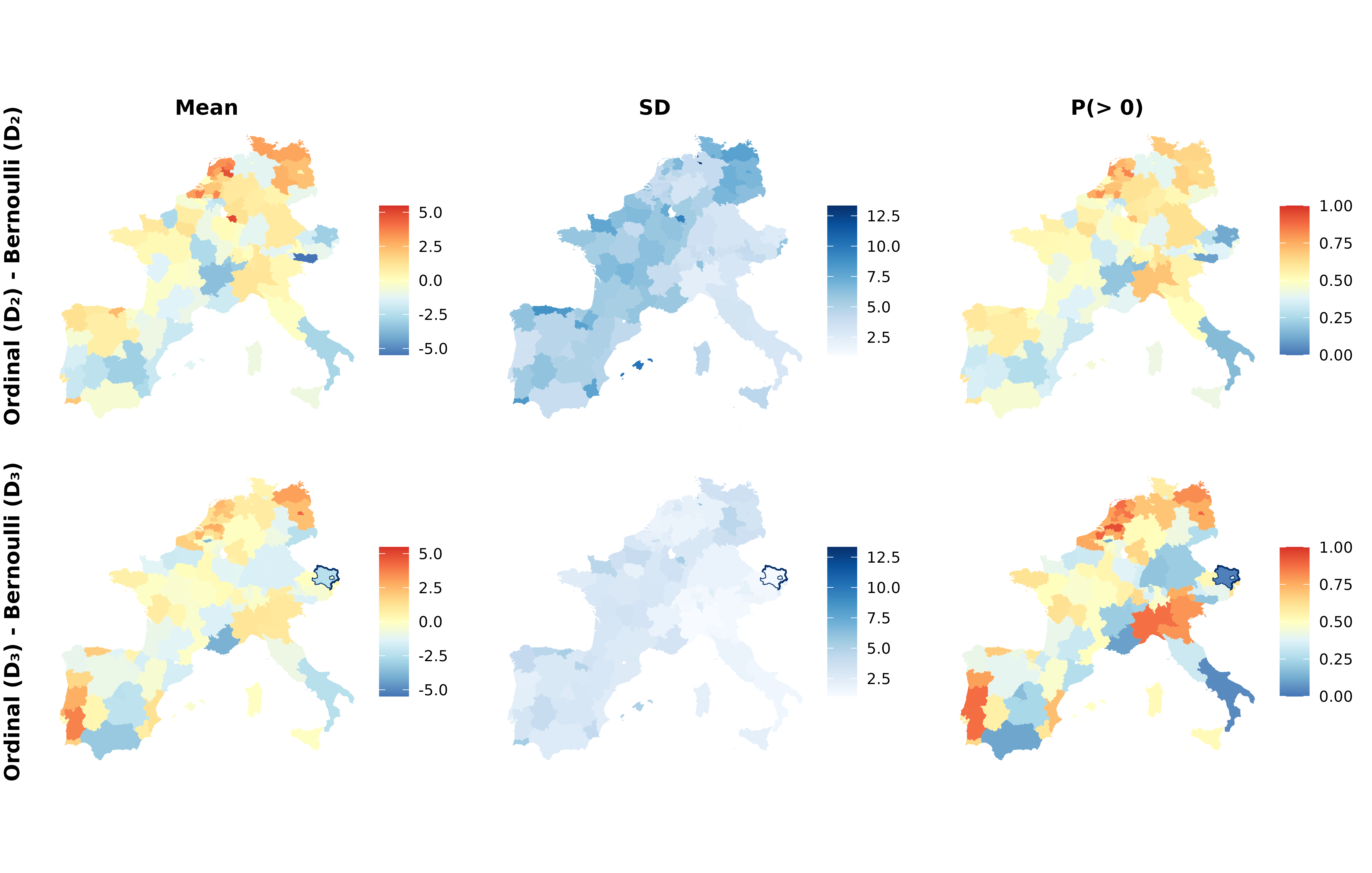}
    \vspace{-0.3cm}
    \caption{Differences between ordinal-model aggregations and Bernoulli poststratified regional prevalences among female respondents. Columns show the posterior mean difference, standard deviation and the probability that the difference is positive. Positive mean values indicate larger ordinal-model estimates. Outlined regions indicate 95\% prediction intervals for the difference excluding zero.}
    \label{fig:differences_woman}
\end{figure}

\begin{figure}[H]
    \hspace{-0.5cm}
    \includegraphics[width=14cm]{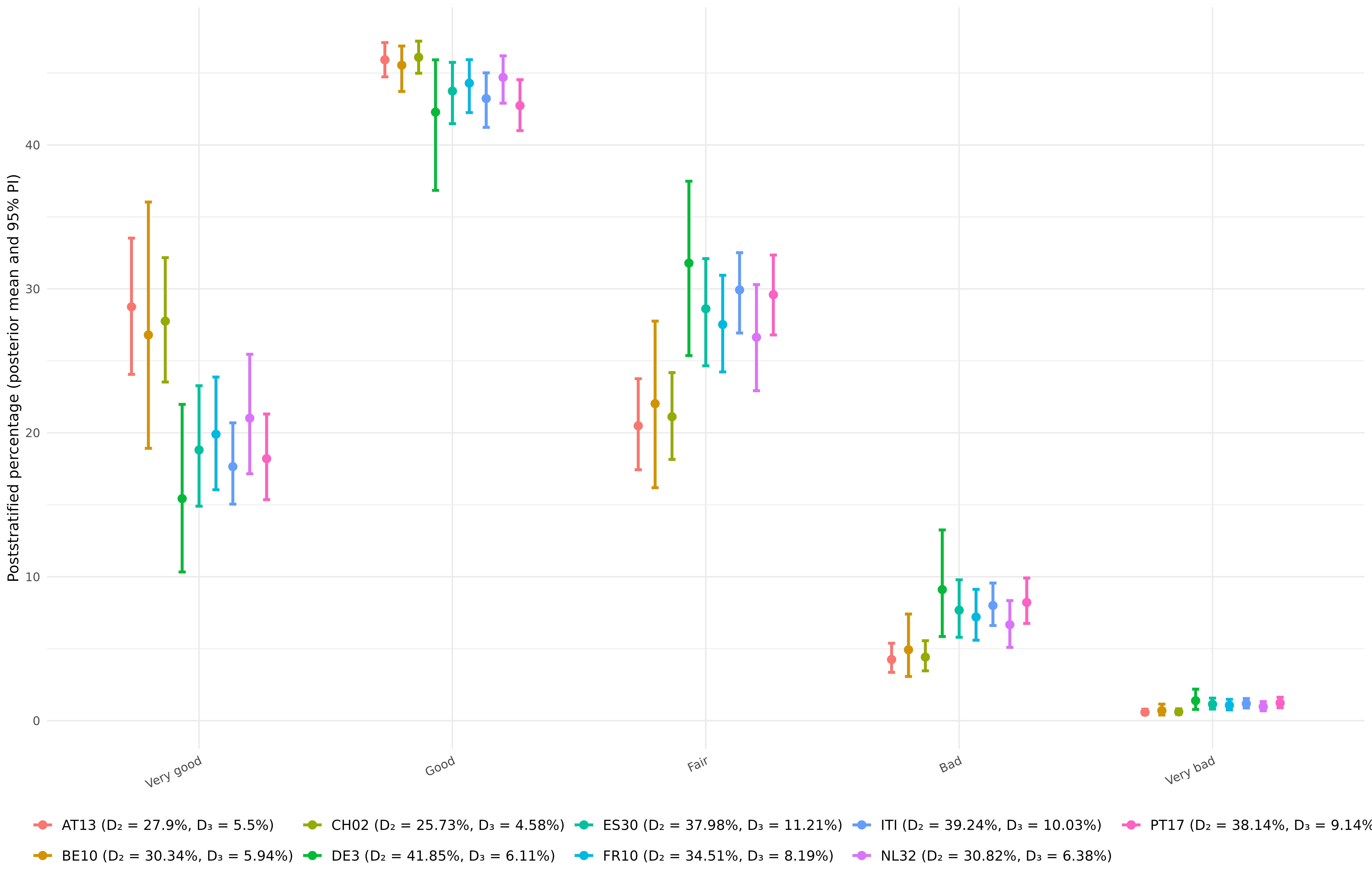}
    \caption{Poststratified ordinal profiles for selected NUTS regions among female respondents. The legend reports posterior means of the poststratified population prevalences for $D_2$ and $D_3$.}
    \label{fig:ordinal_profiles_woman}
\end{figure}

\end{document}